\documentclass[aps,preprint,11pt]{revtex4-1}

\pdfoutput=1
\usepackage{graphicx}
\usepackage{latexsym}
\usepackage[fleqn]{amsmath}
\usepackage{bm}
\usepackage{color}
\usepackage{amssymb}
\usepackage{bm}
\usepackage{ascmac}
\usepackage{bm}

\newcommand{\pa}{\partial}
\newcommand{\la}{\left\langle}
\newcommand{\ra}{\right\rangle}

\newcommand{\Tr}{{\rm Tr}\hspace{0.07cm}}

\begin{document}
\title{Semiclassical optimization of entrainment stability and phase coherence in weakly forced quantum nonlinear oscillators}

\author{Yuzuru Kato}
\email{Corresponding author: kato.y.bg@m.titech.ac.jp}
\affiliation{Department of Systems and Control Engineering,
	Tokyo Institute of Technology, Tokyo 152-8552, Japan}
	
\author{Hiroya Nakao}
\affiliation{Department of Systems and Control Engineering,
	Tokyo Institute of Technology, Tokyo 152-8552, Japan}
\date{\today}

\begin{abstract}
Optimal entrainment of a quantum nonlinear oscillator to a periodically modulated weak harmonic drive is studied in the semiclassical regime.
By using the semiclassical phase reduction theory recently developed for quantum nonlinear oscillators~
[Y. Kato, N. Yamamoto, and H. Nakao,
Semiclassical Phase Reduction Theory for Quantum Synchronization, Phys. Rev. Research 1, 033012 (2019)], two types of optimization problems, one for the stability and the other for the phase coherence of the entrained state, are considered.
The optimal waveforms of the periodic amplitude modulation can be derived by applying the classical optimization methods to the semiclassical phase equation that approximately describes the quantum limit-cycle dynamics.
Using a quantum van der Pol oscillator with squeezing and Kerr effects 
as an example, the performance of optimization is numerically analyzed.
It is shown that the optimized waveform for the entrainment stability yields faster entrainment to the driving signal than the case with a simple sinusoidal waveform, while that for the phase coherence yields little improvement from the sinusoidal case.
These results are explained from the properties of the phase sensitivity function.
\end{abstract}

\maketitle


\section{Introduction}

Synchronization of rhythmic nonlinear systems are widely observed all over the real world, including laser oscillations, mechanical vibrations, and calling frogs \cite{strogatz2004sync,winfree2001geometry, kuramoto1984chemical, pikovsky2001synchronization, ermentrout2010mathematical, nakao2016phase}.
It often plays important functional roles in biological or artificial systems, such as cardiac resynchronization~\cite{ermentrout1984beyond}, phase locked loops in electrical circuits~\cite{best1984phase}, and synchronous power generators~\cite{motter2013spontaneous, dorfler2013synchronization}. 

Recently, experimental studies of synchronization have been performed in micro- and nano-scale nonlinear oscillators~\cite{shim2007synchronized,zhang2012synchronization,
	bagheri2013photonic, kaka2005mutual, weiner2017phase,heimonen2018synchronization}
and theoretical studies of synchronization in the quantum regime have predicted novel features of quantum synchronization~
\cite{
	amitai2017synchronization,
	ludwig2013quantum,weiss2016noise,
	xu2014synchronization,xu2015conditional,
	lee2013quantum,
	walter2014quantum,
	sonar2018squeezing,
	lee2014entanglement,walter2015quantum,
	weiss2017quantum,
	lee2013quantum,lee2014entanglement,hush2015spin,
	roulet2018synchronizing,*roulet2018quantum,nigg2018observing,
	lorch2016genuine,
	witthaut2017classical,
	lorch2017quantum}.
In particular, experimental realization of quantum synchronization is expected in optomechanical oscillators~\cite{bagheri2013photonic, ludwig2013quantum, weiss2016noise, amitai2017synchronization}, oscillators consisting of cooled atomic ensembles~\cite{weiner2017phase, heimonen2018synchronization, xu2014synchronization, xu2015conditional}, and superconducting devices~\cite{nigg2018observing}.
Once realized, quantum synchronization may be applicable in quantum metrology, e.g., improvement of the accuracy of measurements in Ramsey spectroscopy for atomic clocks~\cite{xu2015conditional}.

Nonlinear oscillators possessing a stable limit cycle can be analyzed by using the {\it phase reduction theory}~\cite{kuramoto1984chemical, 
	pikovsky2001synchronization, nakao2016phase}
	 when the forcing given to the oscillator is sufficiently weak.
In the phase reduction theory, multi-dimensional nonlinear dynamical equations describing a limit-cycle oscillator under weak forcing are approximately reduced to a simple one-dimensional phase equation, characterized only by the {\it natural frequency} and {\it phase sensitivity function} (PSF) of the oscillator.
The reduced phase equation enables us to systematically analyze universal dynamical properties of limit-cycle oscillators, such as the entrainment of an oscillator to a weak periodic 
forcing or mutual synchronization of weakly coupled oscillators.

The phase reduction theory has also been used in control and optimization of nonlinear oscillators~\cite{monga2019phase}.
For example, using the reduced phase equations,
minimization of control power for an oscillator~\cite{moehlis2006optimal,zlotnik2012optimal},
maximization of the phase-locking range of an oscillator~\cite{harada2010optimal}, 
maximization of linear stability of an oscillator entrained to a periodic forcing~\cite{zlotnik2013optimal} and of mutual synchronization between two coupled oscillators~\cite{shirasaka2017optimizing, watanabe2019optimization}, 
maximization of phase coherence of noisy oscillators~\cite{pikovsky2015maximizing}, 
and phase-selective entrainment of oscillators~\cite{zlotnik2016phase} have been studied.

\begin{figure} [!t]
	\begin{center}
		\includegraphics[width=0.75\hsize,clip]{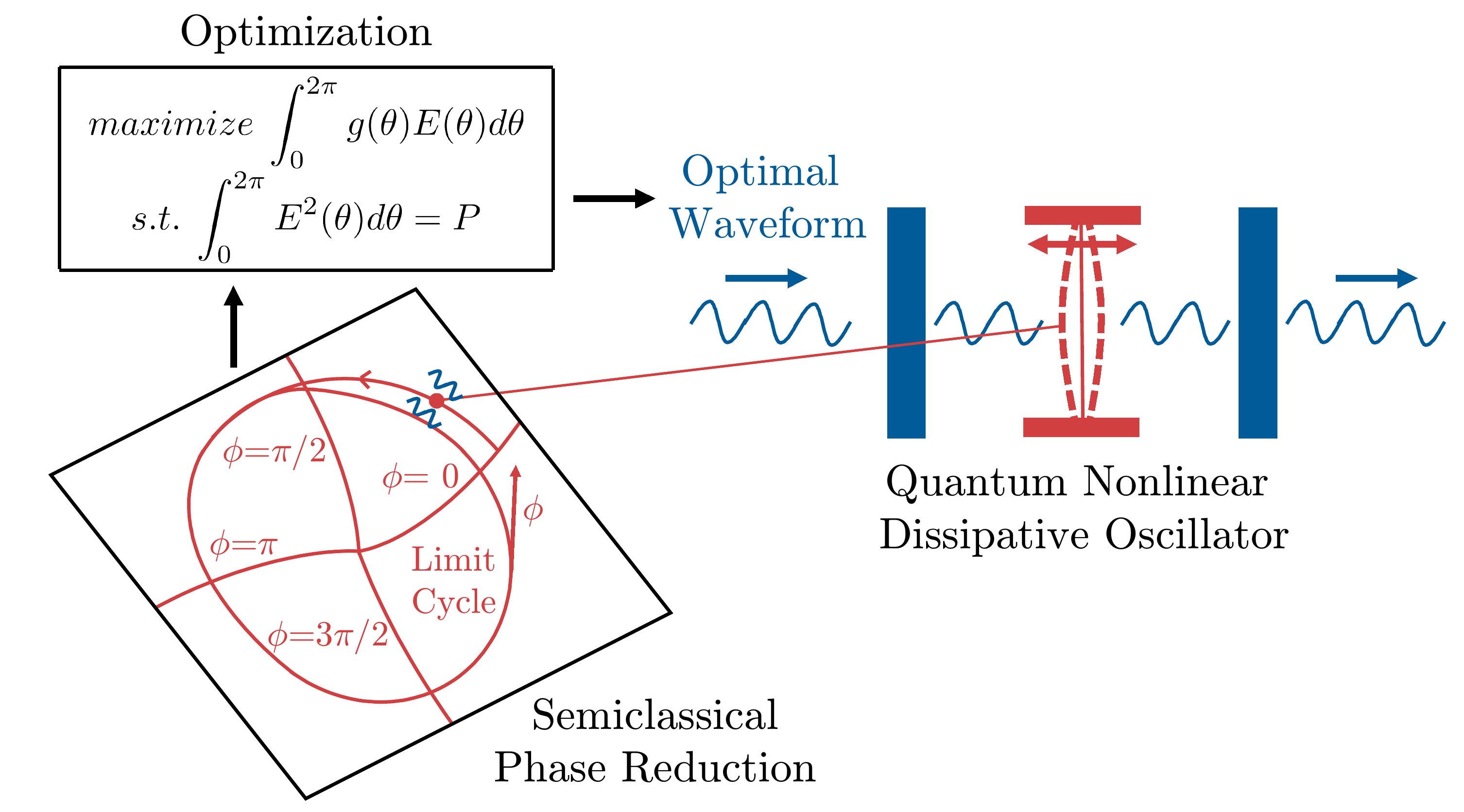}
		\caption{
			A schematic diagram showing optimization of entrainment of a quantum limit-cycle oscillator
			subjected to a periodically modulated harmonic drive. In the semiclassical regime,
			the oscillator can be described by a one-dimensional phase equation. Using the reduced phase 
			equation, we can formulate optimization problems and solve them to derive the optimal waveforms of the periodic amplitude modulation of the harmonic drive.  
			}
		\label{fig_1}
	\end{center}
\end{figure}

Similar to classical nonlinear oscillators, quantum nonlinear oscillators in the semiclassical regime can also be analyzed by using the phase equation.
In Ref.~\cite{hamerly2015optical}, Hamerly and Mabuchi derived a phase equation from the stochastic differential equation (SDE) describing a truncated Wigner function of a quantum limit-cycle oscillator in a free-carrier cavity.
In Ref.~\cite{kato2019semiclassical}, we further developed a phase reduction framework that is applicable to general single-mode quantum nonlinear oscillators.

In this paper, using the semiclassical phase reduction theory~\cite{kato2019semiclassical},
we  optimize entrainment of a quantum nonlinear 
oscillator to a weak harmonic drive with periodic modulation in the semiclassical regime
by employing the optimization methods originally developed for classical oscillators
(see Fig.~\ref{fig_1} for a schematic diagram).
Specifically, we consider two types of optimization problems, i.e., (i) improving entrainment stability~\cite{zlotnik2013optimal} and (ii) enhancing phase coherence~\cite{pikovsky2015maximizing} of the  oscillator.
By using the quantum van der Pol (vdP) oscillator with squeezing and Kerr effects as an example, we illustrate the results of optimization by numerical simulations.

We show that, for the vdP oscillator used in the example, the optimal waveform for the problem (i) leads to larger stability and faster entrainment than the case with the simple sinusoidal waveform, while the optimal waveform for the problem (ii) provides only  tiny enhancement of phase coherence from the sinusoidal case.
 We discuss the difference between the two optimization problems from the properties of the PSF. 
 
This paper is organized as follows. In Sec.~II, we derive a semiclassical phase equation for a weakly perturbed quantum nonlinear oscillator and derive the optimal waveforms for entrainment. In Sec.~III, we illustrate the results of the two optimization methods by numerical simulations and discuss their difference. Sec. IV gives discussion 
and Appendix gives details of calculations.

%
%
\section{Theory}

\subsection{Master equation}
We consider a quantum dissipative system with a single degree of freedom, 
which is interacting with linear and nonlinear reservoirs and
has a stable limit-cycle solution in the classical limit. 
The system is subjected to a weak harmonic drive with a
periodic amplitude modulation of an arbitrary waveform.
Under the Markovian approximation of the reservoirs,
the system obeys a quantum master equation~\cite{gardiner1991quantum, carmichael2007statistical}
\begin{equation}
\label{eq:me}
\dot{\rho}
= -i[H - i \epsilon E(\omega_e t) (a - a^{\dag}), \rho] 
+ \sum_{m=1}^{n} \mathcal{D}[L_{m}]\rho,
\end{equation}
in the rotating coordinate frame of the harmonic drive, where $\rho$ is a density matrix representing the system state, 
$H$ is a system Hamiltonian,
$a$ and $a^{\dag}$ denote annihilation and creation operators ($\dag$ represents Hermitian conjugate), respectively,
$E(\omega_e t)$ is a $2\pi$-periodic scalar function representing the periodic amplitude modulation with frequency $\omega_e$, 
$\epsilon$ is a tiny parameter ($0 < \epsilon \ll 1$) characterizing weakness of the harmonic drive, $n$ is the number of reservoirs, $L_{m}$ 
is the coupling operator between the system and 
$m$th reservoir $(m=1,\ldots,n)$, 
$\mathcal{D}[L]\rho = L \rho L^{\dag} - (\rho L^{\dag} L + L^{\dag} L \rho)/2$ denotes the Lindblad form, and the Planck constant is set as $\hbar = 1$.
It is assumed that the modulation frequency $\omega_e$ is sufficiently close to the natural frequency $\omega$ of the limit cycle in the classical limit.

Using the P representation~\cite{gardiner1991quantum, carmichael2007statistical}, a Fokker-Planck equation (FPE) equivalent to Eq. (\ref{eq:me}) can be derived as
\begin{align}
\label{eq:fpe}
\frac{\pa P(\bm{\alpha}, t)}{\pa t} = \Big[ - \sum_{j=1}^{2} \partial_{j} \{ A_{j}(\bm{\alpha}) + \epsilon E(\omega_e t)  \}
+ \frac{1}{2} \sum_{j=1}^2 \sum_{k=1}^2 \partial_{j}\partial_{k} \{ \epsilon D_{jk}(\bm{\alpha}) \} \Big]P(\bm{\alpha}, t), 
\end{align}
where 
$\bm{\alpha} = (\alpha, \alpha^{*})^\mathsf{T} \in {\mathbb C}^{2 \times 1}$
is a two-dimensional complex vector with $\alpha \in {\mathbb C}$ ($*$ represents complex conjugate and $\mathsf{T}$ represents transpose), $P({\bm \alpha})$ is the P distribution of ${\bm \alpha}$,
$A_j({\bm \alpha})$ is the $j$th components of a complex vector
$\bm{A}(\bm{\alpha}) = (A_{1}(\bm{\alpha})$,
$A^*_{1}(\bm{\alpha}))^\mathsf{T} \in
{\mathbb C}^{2 \times 1} (A_2(\bm{\alpha}) = A^*_{1}(\bm{\alpha}))$
representing the system dynamics, 
$ \epsilon D_{jk}({\bm \alpha})$ is a $(j, k)$-component of a symmetric diffusion matrix $ \epsilon {\bm D}({\bm \alpha}) \in {\mathbb C}^{2 \times 2}$ 
representing quantum fluctuations, and the complex partial derivatives are defined as 
$\partial_1 = \partial / \partial \alpha$ and $\partial_2 = \partial / \partial \alpha^*$.
The drift term $\bm{A}(\bm{\alpha})$
and the diffusion matrix $ \epsilon {\bm D}({\bm \alpha})$
can be calculated from the master equation (\ref{eq:me}) by using the standard operator correspondence for the $P$-representation~\cite{gardiner1991quantum, carmichael2007statistical}.
The
weak harmonic drive with a periodic modulation $\epsilon E(\omega_e t)$ 
and the diffusion matrix $\epsilon{\bm D}({\bm \alpha})$ are 
assumed to be of the same order, $\mathcal{O}(\epsilon)$.

Introducing a complex matrix $ \sqrt{\epsilon} \bm{\beta}(\bm{\alpha}) \in \mathbb{C}^{2 \times 2}$ satisfying
$\epsilon \bm{D}(\bm{\alpha}) = \sqrt{\epsilon} \bm{\beta}(\bm{\alpha})  ( \sqrt{\epsilon} \bm{\beta}(\bm{\alpha}) )^{T}$,
the Ito SDE corresponding to Eq. (\ref{eq:fpe}) for the phase-space variable ${\bm \alpha}(t)$ is obtained as
\begin{align}
\label{eq:ldv}
d\bm{ \alpha}(t) =
\{ {\bm A}({\bm \alpha}(t)) + \epsilon E(\omega_e t)(1, 1)^\mathsf{T} \} dt 
+ 
\sqrt{\epsilon} {\bm \beta}({\bm \alpha}(t)) d{\bm W}(t),
\end{align}
where $\bm{W} = (W_1, W_2)^\mathsf{T} \in \mathbb{R}^{2 \times 1}$ is a vector of independent Wiener processes $W_{i} (i=1,2)$ satisfying 
$\mathbb{E}[{dW_{i}dW_{j}}] = \delta_{ij} dt$
and the explicit form of $\bm{\beta}(\bm{\alpha})$ is given by
\begin{align}
\label{eq:beta}
\bm{\beta}(\bm{\alpha}) &= 
	\begin{pmatrix}
	\sqrt{\frac{\left( R_{12}(\bm{\alpha}) +  R_{11}(\bm{\alpha}) \right)}{2}} e^{i \chi(\bm{\alpha}) / 2}
	&
	-i \sqrt{\frac{\left( R_{12}(\bm{\alpha}) -  R_{11}(\bm{\alpha}) \right)}{2}} e^{i \chi(\bm{\alpha}) / 2}
	\\
	\sqrt{\frac{\left( R_{12}(\bm{\alpha}) +  R_{11}(\bm{\alpha}) \right)}{2}} e^{- i \chi(\bm{\alpha}) / 2}
	&
	i \sqrt{\frac{\left( R_{12}(\bm{\alpha}) -  R_{11}(\bm{\alpha}) \right)}{2}} e^{- i \chi(\bm{\alpha}) / 2}
	\end{pmatrix}
\end{align}
where $R_{11}(\bm{\alpha})  e^{i \chi(\bm{\alpha})} = D_{11}({\bm \alpha})$ and 
$R_{12}(\bm{\alpha}) =  D_{12}({\bm \alpha})$ \cite{kato2019semiclassical}.
In what follows, we only consider the case in which the diffusion matrix is always positive
semidefinite along the limit cycle in the classical limit and derive the phase equation in the two-dimensional phase space of the classical variables~\cite{kato2019semiclassical}.

\subsection{Phase equation and averaging}

As discussed in our previous study~\cite{kato2019semiclassical}, we can derive an approximate SDE for the phase variable of the system from the SDE (\ref{eq:ldv}) in the P representation.
We define a real vector ${\bm X} = (x, p)^\mathsf{T} = ( \mbox{Re}\ \alpha, \mbox{Im}\ \alpha)^\mathsf{T} \in {\mathbb R}^{2 \times 1}$ from the complex vector ${\bm \alpha}$.
Then, the real-valued expression of Eq. (\ref{eq:ldv}) for ${\bm X}$ is
given by an Ito SDE,
\begin{align}
\label{eq:X}
d{\bm{X}}(t) = \{ {\bm{F}}({\bm X}(t)) + \epsilon E(\omega_e t)(1, 0)^\mathsf{T} \} dt 
+ \sqrt{\epsilon} {\bm{G}}({\bm X}(t)) d {\bm W}(t),
\end{align}
where $\bm{F}({\bm X}) \in {\mathbb R}^{2 \times 1}$
and ${\bm{G}}({\bm X}) \in {\mathbb R}^{2 \times 2}$ are real-valued 
representations of the system dynamics ${\bm A}({\bm \alpha}) \in {\mathbb C}^{2 \times 1}$ and noise intensity ${\bm \beta}({\bm \alpha}) 
\in {\mathbb C}^{2 \times 2}$ of Eq. (\ref{eq:ldv}), respectively.

We assume that the system in the classical limit without perturbation and quantum noise, i.e., $\dot{{\bm{X}}} = {\bm{F}}({\bm{X}})$, has an exponentially stable limit-cycle solution ${\bm{X}}_{0}(t) = {\bm{X}}_{0}(t+T)$ with a natural period $T$ and frequency $\omega = 2\pi / T$. 
Following the standard method in the classical phase reduction theory~\cite{winfree2001geometry, kuramoto1984chemical, pikovsky2001synchronization, ermentrout2010mathematical,nakao2016phase}, we can  introduce an asymptotic phase function $\Phi({\bm{X}}) : {\mathbb R}^{2 \times 1} \to [0, 2\pi)$ such that 
$\nabla \Phi({\bm{X}}) \cdot {\bm F}({\bm{X}})  = \omega$
is satisfied in the basin of the limit cycle, where $\nabla \Phi({\bm X}) \in {\mathbb R}^{2 \times 1}$ is the gradient of $\Phi({\bm X})$
\cite{kuramoto1984chemical,
nakao2016phase}.
The phase of a system state ${\bm X}$ is defined as $\phi = \Phi({\bm X})$, which satisfies $\dot{\phi} = \dot{\Phi}({\bm X}) = {\bm F}({\bm X}) \cdot \nabla \Phi({\bm X}) = \omega$ 
($\cdot{}$ represents a scalar product between two vectors).
We represent the system state ${\bm X}$ on the limit cycle as ${\bm X}_0(\phi)$ as a function of the phase $\phi$. Note that an identity $\Phi({\bm X}_0(\phi)) = \phi$ is satisfied 
by the definition of $\Phi({\bm X})$.

Since we assume that the quantum noise and perturbations are sufficiently weak and the deviation of the state ${\bm X}(t)$ from the limit cycle is small, at the lowest-order approximation, we can approximate ${\bm X}(t)$ by ${\bm X}_0(\phi(t))$ and derive a Ito SDE for the phase $\phi$ as
\begin{align}
\label{eq:dphi}
d\phi &= \left\{ \omega + \epsilon \bm{Z}( \phi ) \cdot E(\omega_e t)(1, 0)^\mathsf{T} + \epsilon g(\phi) \right\} dt + \sqrt{\epsilon} \{ \bm{G}(\phi)^\mathsf{T} {\bm Z}(\phi) \} \cdot d\bm{W}.
\end{align}
Here, we introduced the PSF $\bm{Z}(\phi) = \nabla \Phi|_{ {\bm{X} }  = {\bm{X}}_{0}(\phi)} \in {\mathbb R}^{2 \times 1}$ characterizing linear response of the oscillator phase to weak perturbations, a noise intensity matrix $\bm{G}(\phi) = \bm{G}({\bm X}_{0}(\phi))$,
and a function $g(\phi) = \frac{1}{2} \mbox{Tr} \left\{ {\bm G}(\phi)^\mathsf{T} {\bm Y}(\phi) {\bm G}(\phi) \right\}$ where ${\bm Y}(\phi) = \nabla^\mathsf{T} \nabla \Phi|_{{\bm X} = {{\bm X}_0(\phi) }} \in {\mathbb R}^{2 \times 2}$ is a Hessian matrix of the phase function $\Phi({\bm X})$ at ${\bm X} = {\bm X}_0(\phi)$ on the limit cycle.
The PSF \cite{ermentrout2010mathematical} and Hessian \cite{suvak2010quadratic} can be numerically obtained as $2 \pi$-periodic solutions to adjoint-type equations with appropriate constraints. See Ref.~\cite{kato2019semiclassical} for details.

To formulate the optimization problem, we further derive an averaged phase equation from the semiclassical phase equation (\ref{eq:dphi}).
We introduce a phase difference $\psi = \phi - \omega_e t$ between the oscillator and periodic modulation, which is a slow variable obeying 
\begin{align}
\label{eq_psi}
d\psi &= \epsilon \left\{ \Delta_e + Z_x( \psi + \omega_e t) E(\omega_e t) + g(\psi + \omega_e t) \right\} dt
\cr
&+ \sqrt{\epsilon} \{ \bm{G}(\psi + \omega_e t)^\mathsf{T} {\bm Z}(\psi + \omega_e t) \} \cdot d\bm{W},
\end{align}
where $\epsilon\Delta_e = \omega - \omega_e$ and $Z_x$ is the $x$ components of the PSF.
Following the standard averaging procedure~\cite{kuramoto1984chemical}, the small right-hand side of this equation can be averaged over one-period of oscillation via the corresponding FPE~\cite{kato2019semiclassical}, yielding an averaged phase equation
\begin{align}
\label{eq_psiave}
d\psi = \epsilon \left\{ \tilde{\Delta}_e + \Gamma(\psi) \right\} dt
+ \sqrt{\epsilon} \bm{D}_0 \cdot d\bm{W}
\end{align}
which is correct up to $O(\epsilon)$. Here, $\Gamma(\psi)$ is the phase coupling function defined as
\begin{align}
\label{eq:gamma}
\Gamma(\psi) 
= \la Z_x(\psi + \theta) E(\theta) \ra_\theta,
\end{align}
$ \tilde{\Delta}_e = \Delta_e + \la g(\theta) \ra_\theta = 
\omega + \la g(\theta) \ra_\theta - \omega_e = 
\tilde{\omega} - \omega_e$
is the effective detuning of the oscillator frequency from the periodic modulation
($\tilde{\omega} := \omega + \la g(\theta) \ra_{\theta}$ is the 
effective frequency of the oscillator),
$\bm{D}_0 =\la \bm{G}(\theta)^\mathsf{T} {\bm Z}(\theta)  \ra_\theta$,
and the one-period average is denoted as 
$\la \cdot \ra_{\theta} = \frac{1}{2\pi} \int_0^{2\pi} (\cdot) d\theta$.

If the deterministic part of Eq.~(\ref{eq_psiave}) has a stable fixed point at $\psi^*$, the phase of the oscillator can be locked to the periodic amplitude modulation, namely, the phase difference $\psi$ between the oscillator and periodic modulation stays around $\psi^*$ as long as the quantum noise is sufficiently weak.
We consider optimization of the waveform $E$ of the periodic amplitude modulation for (i) improving  entrainment stability and (ii) enhancing phase coherence of the  oscillator.
For the simplicity of the problem, we assume $\tilde{\Delta} = 0$, that is, the frequency of the periodic amplitude modulation is identical with the effective frequency of the system, $\omega_{e} = \tilde{\omega}$.

\subsection{Improvement of entrainment stability}
First, we apply the optimization method of the waveform for stable entrainment, formulated by Zlotnik {\it et al.}~\cite{zlotnik2013optimal} for classical limit-cycle oscillators, to the semiclassical phase equation describing a quantum oscillator.
The entrainment stability is characterized by the linear stability of the phase-locking point $\psi^*$ in the classical limit without noise, which is given by the slope $-\Gamma'(\psi^*)$. The optimization problem is defined as follows:
\begin{align}
\label{eq:opt_obj_fe}
\mbox{maximize}\ -\Gamma^{'}(0),\ \mbox{s.t.}\ \la E^2 (\theta)\ra_{\theta} &= P.
\end{align}
Here, we assume that the phase locking to the periodic modulation occurs at the phase difference $\psi^* = 0$ without losing generality by shifting the origin of the phase.

The solution to this problem maximizes the linear stability $-\Gamma'(0)$ of the fixed point $\psi^*=0$ of the deterministic part of Eq.~(\ref{eq_psiave}).
Maximization of the linear stability minimizes the convergence time to the fixed point, resulting in faster entrainment of the oscillator to the driving signal when the noise is absent.
This problem is solved under the condition that the control power $\langle E^2(\theta) \rangle_\theta$ is fixed to $P$, where $P$ is assumed to be sufficiently small.
As derived in Appendix, the optimal waveform for Eq.~(\ref{eq:opt_obj_fe}) is explicitly given by
\begin{align}
\label{eq:opt_ipt_fe}
E(\theta) =
- \sqrt{\frac{P}{ \la Z_x'(\theta) ^2 \ra_\theta }} Z_x'(\theta),
\end{align}
which is proportional to the differential of the $x$ component $Z_x(\theta)$ of the PSF. 

\subsection{Enhancement of phase coherence}
Next, we apply the optimization method of the waveform for enhancement of phase coherence in the weak noise limit, which was formulated by Pikovsky~\cite{pikovsky2015maximizing} for classical noisy limit-cycle oscillators, to the semiclassical phase equation describing a quantum oscillator.
In the weak noise limit, the phase coherence is characterized by the depth $v(\psi_{max}) - v(\psi^*)$ of the potential $v(\psi) = \int^{\psi} \{ - \Gamma(\theta) \} d\theta$ of the deterministic part of Eq.~(\ref{eq_psiave}), where $\psi_{max}$ and $\psi^*$ give the maximum and minimum of the potential $v(\psi)$, respectively (we assume that $\psi^*$ corresponds to the potential minimum, i.e., we focus on the most stable fixed point if there are multiple stable fixed points).
In this case, the optimization problem is defined as follows:
\begin{align}
\label{eq:opt_obj_co}
\mbox{maximize}\ \int_{\psi^*}^{\psi_{max}} \{ - \Gamma(\psi) \} d\psi, ~ \mbox{s.t.}~\la E^2 (\theta)\ra_{\theta} &= P.
\end{align}

The solution to this optimization problem maximizes the depth of the potential $v(\psi)$ at the phase-locked point, thereby minimizing the escape rate of noise-induced phase slipping and maximizing the phase coherence of the oscillator under sufficiently weak noise, as discussed in Ref.~\cite{pikovsky2015maximizing} for the classical case.
As in the previous problem, this optimization problem is solved under the condition that the control power $\la E^2 (\theta)\ra_{\theta}$ is fixed to $P$.

In what follows, 
we introduce $\Delta_{\psi} = \psi_{max} - \psi^*$ and assume $\psi^* = 0$ without loss of generality. Then, the optimal waveform is obtained as (see Appendix for the derivation)
\begin{align}
\label{eq:opt_ipt_co}
E(\theta) =
-\sqrt{\frac{P}{ \la ( \int_{\theta}^{\theta + \Delta_{\psi}} Z_x(\phi) d\phi )^2 \ra_\theta }} \int_{\theta}^{\theta + \Delta_{\psi}} Z_x(\phi) d\phi,
\end{align}
which is proportional to the integral of the $x$ component $Z_x(\phi)$ of the PSF, in contrast to the previous case in which the optimal waveform is proportional to the differential of $Z_x(\phi)$.


\section{Results}

\subsection{Quantum van der Pol oscillator}
As an example, we consider a quantum vdP oscillator with squeezing and Kerr effects subjected to a periodically modulated harmonic drive.
In our previous study~\cite{kato2019semiclassical}, we have analyzed entrainment of a vdP oscillator with only a squeezing effect to a purely sinusoidal periodic modulation; in this study, we seek optimal waveforms of the periodic modulation for a vdP oscillator with both squeezing and Kerr effects.
We use QuTiP numerical toolbox for direct numerical simulations of the master equation~\cite{johansson2012qutip,*johansson2013qutip}.

We assume that the harmonic drive is sufficiently weak and treat it as a perturbation, while the squeezing and  Kerr effects are both relatively 
strong and cannot be treated as perturbations.
The frequencies of the oscillator, harmonic drive, and pump beam of squeezing are denoted by $\omega_{0}$, $\omega_{d}$, and $\omega_{sq}$, respectively. 
We consider the case in which the squeezing is generated by a degenerate parametric amplifier and we set $\omega_{sq} = 2\omega_{d}$.

In the rotating coordinate frame of frequency $\omega_{d}$, the master equation for the quantum vdP oscillator is given by
~\cite{kato2019semiclassical, lorch2016genuine}
\begin{align}
\label{eq:qvdp_me}
\dot{\rho} 
= 
-i [  - \Delta a^{\dag}a + K a^{\dag 2} a^2 -  i E(\omega_e t) (a - a^{\dag}) +
 i \eta ( a^2 e^{-i \theta} - a^{\dag 2} e^{ i \theta}  )
,\rho ]
+ \gamma_{1} \mathcal{D}[a^{\dag}]\rho + \gamma_{2}\mathcal{D}[a^{2}]\rho,
\end{align}
where $\Delta = \omega_{d} - \omega_{0}$ is the frequency detuning  
of the harmonic drive from the oscillator, 
$K$ is the Kerr parameter,
$E(\omega_e t)$ is the periodic amplitude modulation
of the harmonic drive, $\eta e^{ i \theta}$ 
is the squeezing parameter, 
$\gamma_{1}$ and $\gamma_{2}$ are the decay rates for 
negative damping and nonlinear damping, respectively.

We assume $\gamma_{2}$ to be sufficiently small, for which the semiclassical approximation is valid, and represent $\gamma_2$ as $\gamma_{2} = \epsilon \gamma_{1} \gamma_{2}^{'}$
with a dimensionless parameter $\gamma_{2}^{'}$ of $\mathcal{O}(1)$.
As discussed in Ref.~\cite{kato2019semiclassical}, to rescale the size of the limit cycle to be $O(1)$, we introduce a rescaled annihilation operator $a'$, classical variable $\alpha'$,
and rescaled parameters $\Delta = \gamma_{1} \Delta{'}, K = \epsilon \gamma_{1} K^{'}, E(\omega_e t) = \sqrt{\epsilon} \gamma_{1} E{'}(\omega_e t), \eta = \gamma_{1} \eta{'}$, where $\Delta', K', E', \eta'$ are dimensionless parameters of $\mathcal{O}(1)$.
We also rescale the time and frequency of the periodic modulation as $t' = \gamma_{1} t$ and $\omega_e = \gamma_1 \omega'_e$, respectively. The FPE for the P distribution corresponding to Eq.~(\ref{eq:qvdp_me}) is then given by
\begin{align}
\label{eq:rescaled_fpe}
\hspace*{-2em}
\frac{\pa P(\bm{\alpha'}, t')}{\pa t'} = \Big[ - \sum_{j=1}^{%
	2} \partial'_{j} \{ A_{j}(\bm{\alpha'}) + \epsilon 
E'(\omega'_e t') \} 
+ \frac{1}{2} \sum_{j=1}^2 \sum_{k=1}^2 \partial'_{j}\partial'_{k} \{ \epsilon D_{jk}(\bm{\alpha'}) \} \Big]P(\bm{\alpha'}, t'), 
\end{align}
where $\bm{\alpha'} = (\alpha', \alpha'^{*}) = \sqrt{\epsilon}( \alpha, \alpha^{*}) $,
$\partial'_1 = \partial / \partial \alpha'$, $\partial'_2 = \partial / \partial \alpha'^*$, 
\begin{align}
\label{eq:qvdp_drift}
&\bm{A}(\bm{\alpha'})  =
	\left( \begin{matrix}
	\left(\frac{1}{2} + i \Delta' \right) \alpha'   
	- (\gamma'_{2}  + 2  K'  i ) \alpha'^{*} \alpha'^{2} 
	- 2 \eta' e^{i \theta}\alpha'^{*}
	\\
	\left(\frac{1}{2} - i \Delta' \right) \alpha'^{*}   
	- (\gamma'_{2}  - 2  K'  i ) \alpha'\alpha'^{*2} 
	- 2 \eta' e^{- i \theta}\alpha'
	\\
	\end{matrix} \right),
\end{align}
and
\begin{align}
\label{eq:qvdp_diffusion}
& \bm{D}(\bm{\alpha'}) =
	\left( \begin{matrix}
	- (( \gamma'_{2} + 2 K' i) \alpha'^{2}  + 2 \eta' e^{ i \theta} )   & 1 \\
	1 & -(( \gamma'_{2} - 2 K' i)  \alpha'^{*2}  + 2  \eta' e^{ - i \theta}) \\
	\end{matrix} \right).
\end{align}

The real-valued vector 
$\bm{X} = (x', p')^\mathsf{T} = (\mbox{Re}~\alpha', \mbox{Im}~\alpha')^\mathsf{T}$
of Eq. (\ref{eq:X}) after rescaling is
\begin{align}
\label{eq:qvdp_X}
&d \bm{X} =
\cr
&
	\begin{pmatrix}
	\frac{1}{2}x'  - \Delta' p'  
	- (\gamma'_{2} x' - 2 K' p') (x'^{2} + p'^{2}) 
	+ \epsilon E'(\omega'_e t') - 2 \eta' ( x' \cos \theta + p' \sin \theta)  
	\cr
	\frac{1}{2}p' + \Delta' x'  
	- (\gamma'_{2} p' + 2 K' x') (x'^{2} + p'^{2}) 
	+ 2 \eta' ( p' \cos \theta - x' \sin \theta) 
	\end{pmatrix}
dt 
\cr
&
+\sqrt{\epsilon} \bm{G}(\bm{X}) d\bm{W}',
\end{align}
where $d\bm{W}' = \sqrt{\gamma_{1}}d\bm{W}$ and the noise intensity matrix is explicitly given by 
\begin{align}
\label{eq:noiseintensity}
\bm{G}(\bm{X}) & =
\left( \begin{matrix}
\sqrt{\frac{\left( 1 +  R'_{1} \right)}{2}} 
\cos \frac{\chi'_1}{2} & 
\sqrt{\frac{\left( 1 -  R'_{1} \right)}{2}} 
\sin \frac{\chi'_1}{2} \\
\sqrt{\frac{\left( 1 +  R'_{1} \right)}{2}} 
\sin \frac{\chi'_1}{2} & 
- \sqrt{\frac{\left( 1 -  R'_{1} \right)}{2}}  
\cos \frac{\chi'_1}{2} \\
\end{matrix} \right)
\end{align}
with $R'_{1}e^{i \chi'_1} = - (( \gamma'_{2} + 2 K' i) \alpha'^{2}  + 2 \eta' e^{ i \theta} )$.
The deterministic part of Eq.~(\ref{eq:qvdp_X}) without the harmonic drive~($E'=0$) gives an asymmetric limit cycle when $\eta' > 0$ and cannot be solved analytically.
Hence, we numerically obtain the limit cycle ${\bm X}_0(\phi)$ and evaluate the PSF ${\bm Z}(\phi)$, Hessian matrix ${\bm Y}(\phi)$, and noise intensity ${\bm G}(\phi)$.
We then use these quantities to derive the optimal waveforms.

We consider two parameter sets, which correspond to (i) a limit cycle with asymmetry due to the effect of squeezing, $(\Delta, \gamma_{2}, \eta e^{i \theta}, K)/\gamma_{1} = (0.575, 0.05, 0.2, 0)$, and (ii) a limit cycle with asymmetry due to squeezing and Kerr effects, $(\Delta, \gamma_{2}, \eta e^{i \theta}, K)/\gamma_{1} = (0, 0.05, 0.15, 0.03)$.
Note that we use parameter sets for which the limit cycles in the classical limits are asymmetric for the evaluation of the optimization methods. This is because the optimal waveform is given by a trivial sinusoidal function when the limit cycle is symmetric and the $x$ component of the PSF has 
a sinusoidal form (see Appendix).
We set the control power as $P = \sqrt{0.2}$ and compare the results for optimal waveforms with those for the simple sinusoidal waveform.

\begin{figure} [!t]
	\begin{center}
		\includegraphics[width=0.5\hsize,clip]{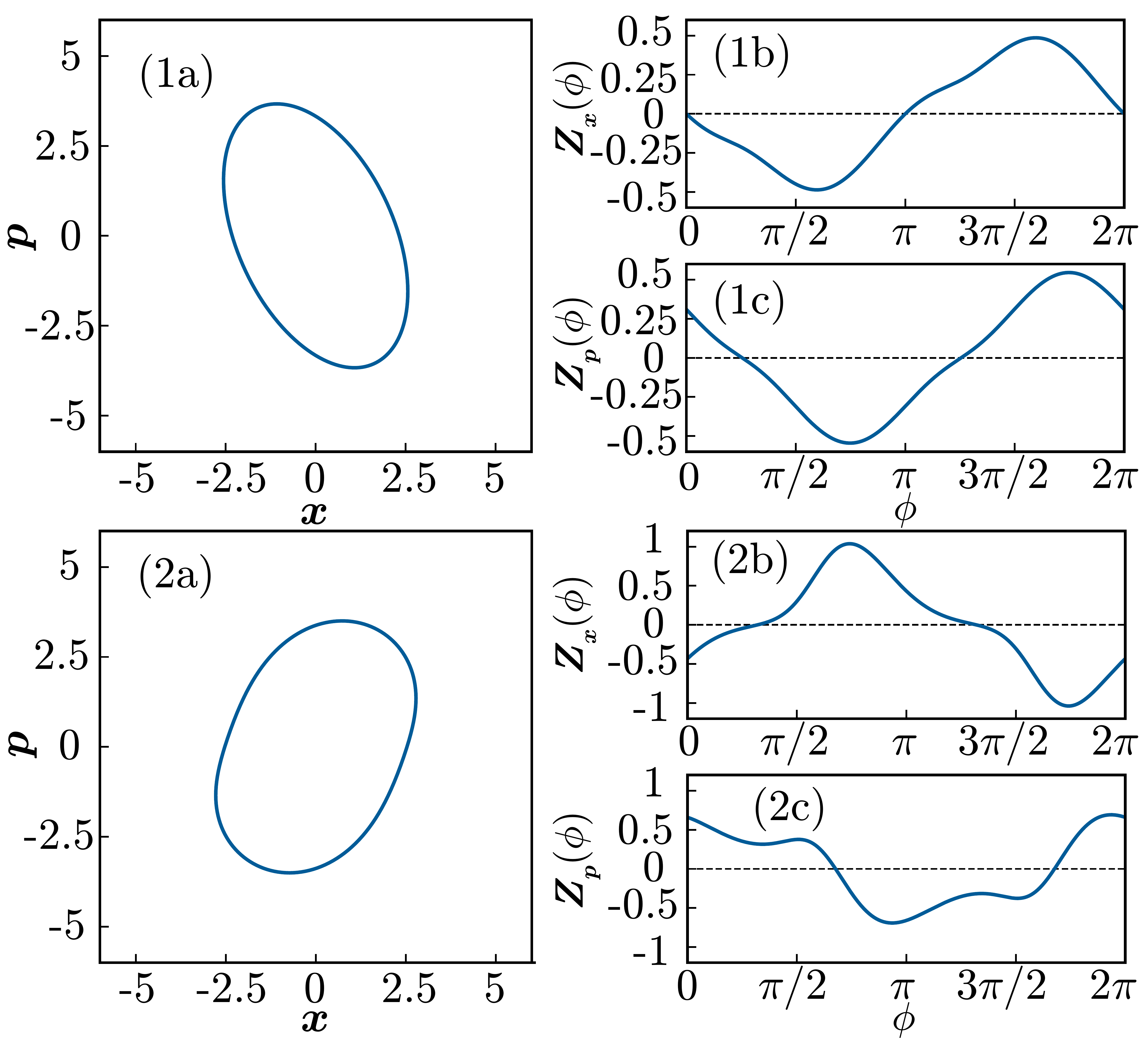}
		\caption{
			Limit cycles and phase sensitivity functions of a quantum van der Pol oscillator with only the
			squeezing effect (1a, 1b, 1c) and with both squeezing and Kerr effects (2a, 2b, 2c).
			(1a,2a): Limit cycle ${\bm X}_0(\phi)$ in the classical limit.
			\; 
			(1b,2b): $x$ component $Z_x(\phi)$ of the PSF ${\bm Z}(\phi)$.
			\;
			(1c,2c): $p$ component $Z_p(\phi)$ of the PSF ${\bm Z}(\phi)$.
			Note that the figures are drawn using $x$ and $p$ before rescaling.
}
		\label{fig_2}
	\end{center}
\end{figure}

Figures~\ref{fig_2} (1a-1c) and (2a-2c) show the limit cycles and PSFs
in the classical limit for the cases (i) and (ii), respectively.
The natural and effective frequencies of the oscillator are $(\omega,\tilde{\omega}) = (0.413, 0.407)$ in the case (i) and $(\omega,\tilde{\omega}) = (0.510,0.451)$ in the case (ii), respectively.
In the case (i), the drift coefficient of the phase variable is positive when the oscillator rotates counterclockwise and the origin of the phase $\phi=0$ 
is chosen as the intersection of the limit cycle and the ${x}'$ axis with ${x}'>0$.
In the case (ii), the drift coefficient of the phase variable is positive when the oscillator rotates clockwise and the origin of the phase $\phi=0$ is chosen as the intersection of the limit cycle and the ${x}'$ axis with ${x}'<0$.

\subsection{Improvement of entrainment stability}

To evaluate the performance of the optimal waveform for the entrainment stability,
we use half the square of the Bures distance
$F_q(\rho_1, \rho_2) = 1 - \Tr [ \sqrt{ \sqrt{\rho_{2}} \rho_{1}  \sqrt{\rho_{2}}}]$
obtained by direct numerical simulations of the master equation (\ref{eq:qvdp_me})
and the corresponding classical distance
$F_c(P_1(\psi), P_2(\psi )) = 1- \la \sqrt{P_1(\psi) P_2(\psi)} \ra_{\psi}$
for the probability distributions of 
the phase variable~\cite{bures1969extension}
obtained from the reduced phase equation (\ref{eq:dphi}).
We consider the distance between the system states at $t$ and $t+T_e$ 
with $T_e = 2 \pi/\omega_e$ (i.e., one period later), and use $F_q(\rho_t, \rho_{t + T_e})$ and $F_c(P_t(\psi), P_{t+T_e}(\psi))$ to measure the performance, since $F_q(\rho_t, \rho_{t + T_e})$ and $F_c(P_t(\psi), P_{t+T_e}(\psi))$ converge to zero when the system converges to a periodic steady (cyclo-stationary) state with period $T_e$.

\begin{figure} [!t]
	\begin{center}
		\includegraphics[width=1\hsize,clip]{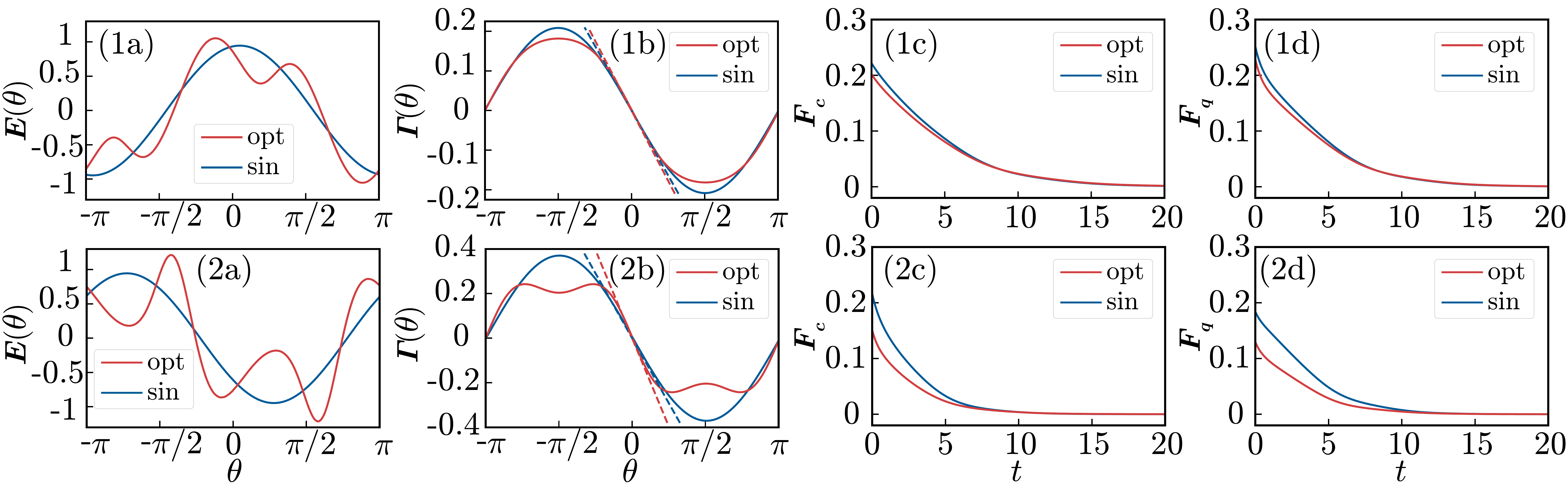}
		\caption{
			Results of optimization for the entrainment stability in the case (i) (1a-1d) and case (ii) (2a-2d).
			Red lines show the results for the optimal waveform, and 
			blue lines show the results for the sinusoidal waveform.
			(1a,2a): Optimal waveform $E$ of the periodic amplitude modulation.
			(1b,2b): Interaction function $\Gamma$.
			(1c,2c): Classical distance $F_c$.
			(1d,2d): Quantum distance $F_q$.
		}
		\label{fig_3}
	\end{center}
\end{figure}

To eliminate the dependence on the initial phase $\theta_0$ of the input,
we calculate $F_{c,q}^{\theta_0}$ by using an input signal $E(\omega_e t + \theta_0)$, average it over $0 \leq \theta_0 < 2\pi$ to obtain $\la F_{c,q}^{\theta_0} \ra_{\theta_0}$,
and use this as the measure for evaluating the entrainment of the oscillator.
We set the initial state of the density matrix as the steady state of Eq.~(\ref{eq:qvdp_me}) 
without the periodically modulated harmonic drive (E = 0), and the initial state
of the corresponding phase distribution as a uniform distribution $P(\psi) = 1 / (2\pi)$.
Figures~\ref{fig_3}(1a-1d) and ~\ref{fig_3}(2a-2d) show the results for the cases (i) and (ii), respectively, where the optimal waveforms of $E$ are plotted in Figs.~\ref{fig_3}(1a, 2a), the phase-coupling functions $\Gamma$ are plotted in Figs.~\ref{fig_3}(1b, 2b), the classical distances $F_c$ are plotted in Figs.~\ref{fig_3}(1c, 2c), and the quantum distance $F_q$ are plotted in Figs.~\ref{fig_3}(1d, 2d).

In the case (i), the linear stability of the entrained state is given by $-\Gamma'_{opt}(0) = 0.226$ in 
the optimized case, which is higher than $-\Gamma'_{sin}(0) = 0.208$ in the sinusoidal case by a factor $\Gamma_{opt}'(0)/\Gamma_{sin}'(0) = 1.083$.
As a result, faster entrainment to the entrained state can be observed in both Figs.~\ref{fig_3}(1c) and ~\ref{fig_3}(1d) in the optimized cases.
In the case (ii), the linear stability is given by $-\Gamma'_{opt}(0) = 0.503$ in the optimized case, which is higher than $-\Gamma'_{sin}(0) = 0.371$ in the sinusoidal case by a factor $\Gamma_{opt}'(0)/\Gamma_{sin}'(0) = 1.358$. 
Faster entrainment to the entrained state can also be confirmed from Figs.~\ref{fig_3}(2c) and ~\ref{fig_3}(2d), where both $F_{c}$ and $F_{q}$ converge faster in the optimized cases.

Note that larger improvement factor is attained in the case (ii) than in the case (i), which results from stronger anharmonicity of the PSF in the case (ii) than in the case (i). This point will be discussed in 
Sec.~III~D.

\subsection{Enhancement of phase coherence}
To evaluate the performance of the optimal waveform for the phase coherence, we use the averaged maximum value of the Wigner function $\la \max W^{\psi} \ra_{\psi}$, where $W^{\psi}$ is the Wigner distribution of the density matrix $\rho$ at phase $\psi$ of the periodic steady state obtained by direct numerical simulations
of the master equation (\ref{eq:qvdp_me}).
We also use the averaged maximum value for the corresponding probability distribution of the phase variable $\la \max P^{\psi} \ra_{\psi}$, where $P^{\psi}$ is the probability distribution at phase $\psi$ of the periodic steady state obtained from the reduced phase equation (\ref{eq:dphi}).

\begin{figure} [!t]
	\begin{center}
		\includegraphics[width=0.6\hsize,clip]{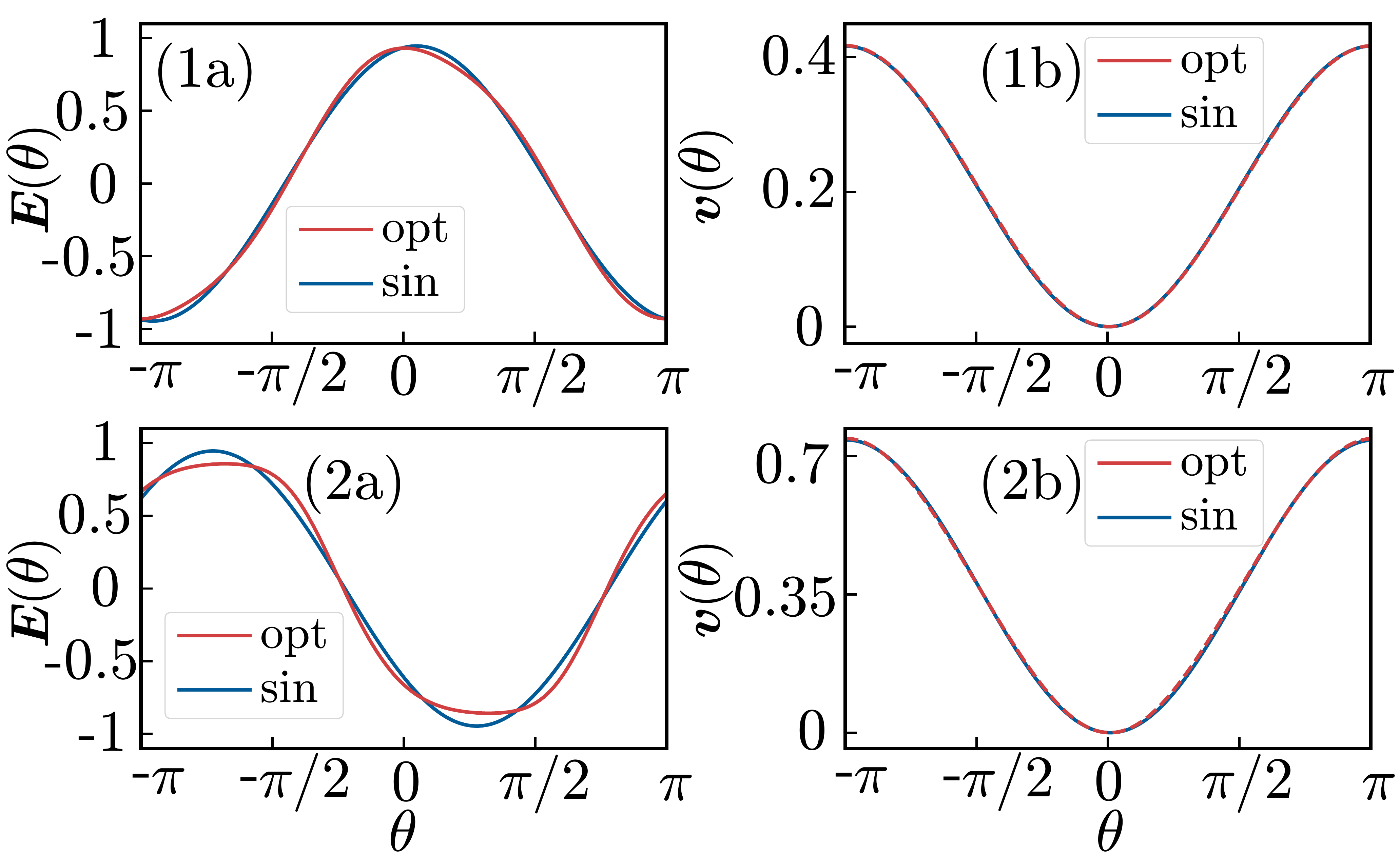}
		\caption{
			Results for enhancement of phase coherence 
			in the case (i) (1a, 1b) and case (ii) (2a, 2b).
			Red lines show the results for the optimal waveform, and
			blue lines show the results for the sinusoidal waveform, respectively.
			(1a,1b): Optimal waveform $E$ of the periodic amplitude modulation.
			(2a,2b): Potential $v$ of the phase difference.
			}
		\label{fig_4}
	\end{center}
\end{figure}

Figure~\ref{fig_4}(1a) and ~\ref{fig_4}(2a) show the optimal waveforms of $E$, and Fig.~\ref{fig_4}(1b) and ~\ref{fig_4}(2b) show the potential $v$ of the phase difference. 
In the case (i), the maximum value of the potential $v$ is given by $ v_{opt}(\Delta_{\psi}) = 0.4172$ in the optimized case, which is slightly higher than $ v_{sin}(\Delta_{\psi}) = 0.4167$ in the sinusoidal case by a factor $v_{opt}(\Delta_{\psi})/v_{sin}(\Delta_{\psi}) = 1.001$.
Accordingly, we obtain a tiny enhancement of phase coherence from the averaged maximum values of both the Wigner distribution of the quantum system $\la \mbox{max} W^{\psi}_{opt} \ra_{\psi}/\la \mbox{max} W^{\psi}_{sin} \ra_{\psi} = 1.0028$ and the corresponding probability distribution of the classical phase variable
$\la \mbox{max} P^{\psi}_{opt} \ra_{\psi}/\la \mbox{max} P^{\psi}_{sin} \ra_{\psi} = 1.0076$, although it is difficult to see the difference from Fig.~\ref{fig_4}(1b) itself.

In the case (ii), the maximum value of the potential $v$ is given by $ v_{opt}(\Delta_{\psi}) = 0.7447$ in
the optimized case, which is also slightly higher than $ v_{sin}(\Delta_{\psi}) = 0.7411$ in the sinusoidal case by $v_{opt}(\Delta_{\psi})/v_{sin}(\Delta_{\psi}) = 1.005$.
We obtain a tiny enhancement of phase coherence from both the averaged maximum values of the Wigner function of the quantum system $\la \mbox{max} W^{\psi}_{opt} \ra_{\psi}/\la \mbox{max} W^{\psi}_{sin} \ra_{\psi} = 1.0063$ and the corresponding probability distribution of the classical phase variable $\la \mbox{max} P^{\psi}_{opt} \ra_{\psi} / $ $\la \mbox{max} P^{\psi}_{sin} \ra_{\psi} = 1.0143$.
 
For the vdP oscillator used here, only tiny enhancements in the phase coherence 
can be observed in both case (i) and case (ii). This is because the PSF does not have strong high-harmonic components in both cases (see Fig.~\ref{fig_5}). It should also be noted that the improvement factor in the case (ii) is larger than in case (i), which results from stronger anharmonicity of the PSF in the case (ii) than in the case (i). We discuss these points in Sec.~III~D.

\subsection{Comparison of two optimization problems}

In Sec.~III~B, we could observe that the optimized waveforms yield clearly faster convergence to the entrained state than the sinusoidal waveform, indicating improvements in the stability of the entrained state, while in Sec.~III~C, we could observe only tiny enhancements in the phase coherence from the sinusoidal case.
This difference between the two optimization problems can be explained from the general expressions for the optimized waveforms.

The optimal waveform for the entrainment stability is proportional to the differential of the $x$ component $Z_x$ of the PSF as can be seen from Eq.~(\ref{eq:opt_ipt_fe}), while that for the phase coherence is proportional to the integral of $Z_x$ as in Eq.~(\ref{eq:opt_ipt_co}). 
Because the PSF is a $2\pi$-periodic function, $Z_x$ can be expanded in a Fourier series as
\begin{align}
Z_x(\theta) = \sum_{n=-\infty}^{\infty} Z_n \exp [ i n \theta],
\end{align}
where $Z_n$ ($n=0, \pm1, \pm2, \cdots$) are the Fourier coefficients.
The differential of $Z_x(\theta)$ can then be expressed as
\begin{align}
Z'_x(\theta) = \sum_{n=-\infty}^{\infty} i n Z _n\exp [ i n \theta ],
\end{align}
and the integral of $Z_x(\theta)$ can be expressed as
\begin{align}
\int_{\theta}^{\theta + \Delta_{\psi}} Z_x(\theta) d\psi = 
\sum_{n=-\infty (n \neq 0)}^{\infty} \frac{ Z _n ( \exp [ i n ( \theta + \Delta_{\psi}) ]  - \exp [ i n \theta]) }{in},
\end{align}
where $n = 0$ is omitted from the sum to avoid vanishing denominator without changing the result.
Thus, the deviation of the differential $Z'_x(\theta)$ from the sinusoidal function is larger because the $n$th Fourier component is multiplied by $n$, while the deviation of the integral $\int_{\theta}^{\theta + \Delta_{\psi}} Z_x(\psi) d\psi$ from the sinusoidal function is smaller because the $n$th Fourier component is divided by $n$.
This explains the difference in the performance of the two optimization problems, namely, why we observed considerable improvement in the entrainment stability while only tiny improvement in the phase coherence from the simple sinusoidal waveform.

\begin{figure} [!t]
	\begin{center}
		\includegraphics[width=0.5 \hsize,clip]{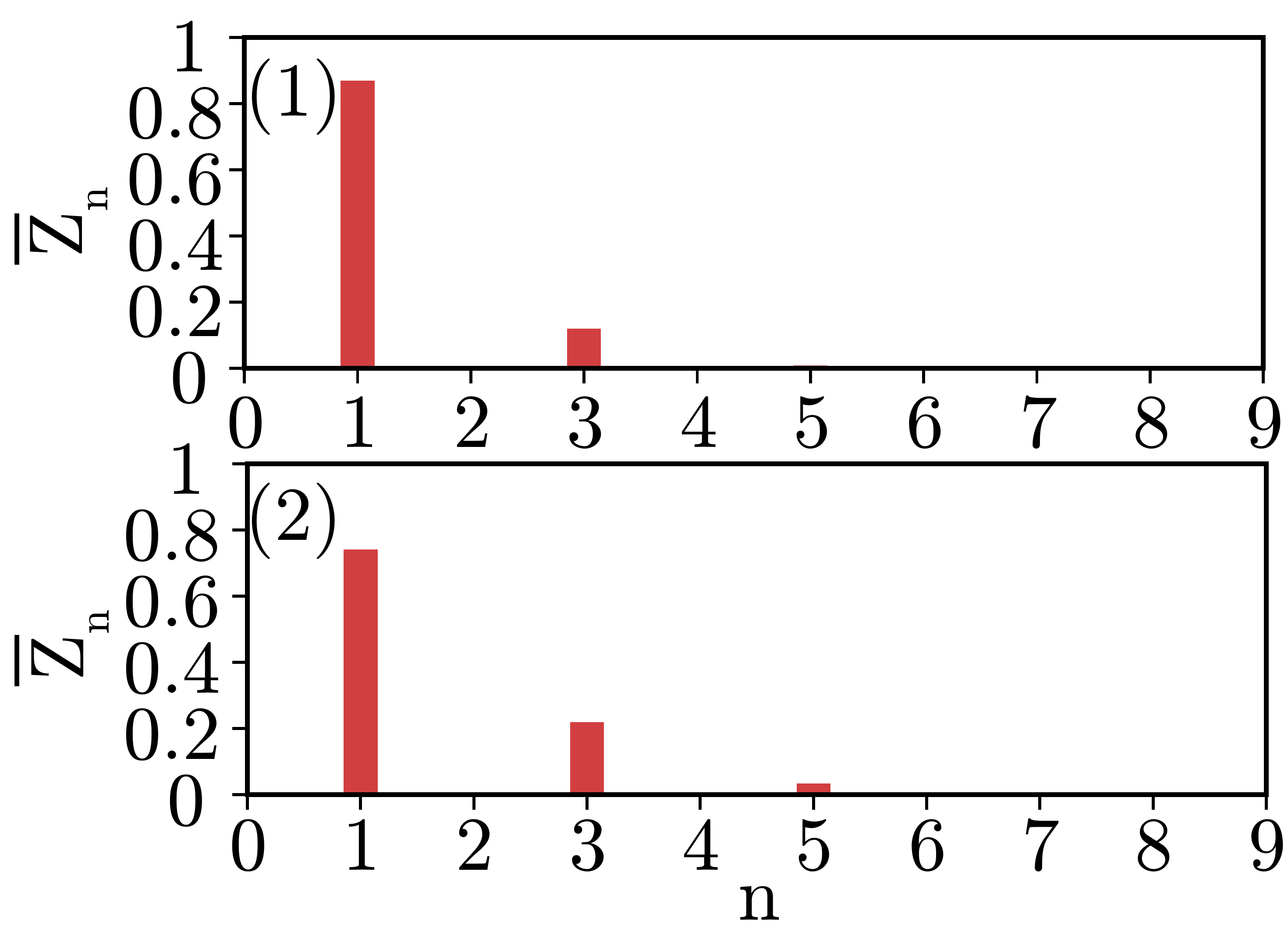}
		\caption{
			Normalized absolute value of Fourier components
			$\bar{Z_n} = |Z_n|/\sum_{n=0}^{\infty}|Z_n|$ in the cases (i) (top) and (ii) (bottom).
			In case (i)
			$(\bar{Z_0}, \bar{Z_1}, \bar{Z_2}, \bar{Z_3},\bar{Z_4},\bar{Z_5},\bar{Z_6},\bar{Z_7},\bar{Z_8},\bar{Z_9}) 
			= (0, 0.87, 0, 0.12, 0, 0.009, 0, 0.001, 0, 0)$,
			and in case (ii)
			$(\bar{Z_0}, \bar{Z_1}, \bar{Z_2}, \bar{Z_3},\bar{Z_4},\bar{Z_5},\bar{Z_6},\bar{Z_7},\bar{Z_8},\bar{Z_9}) 
			= (0, 0.741, 0, 0.219, 0, 0.034, 0, 0.005, 0, 0.001)$,
			respectively.
		}
		\label{fig_5}
	\end{center}
\end{figure}

From the above expressions, we also find that the deviations of $Z'_x(\theta)$ and $\int_{\theta}^{\theta + \Delta_{\psi}} Z_x(\theta) d\psi$ from the sinusoidal function are more pronounced when the PSF possesses stronger high-frequency components.  %
Figures~\ref{fig_5}(1a,1b) and ~\ref{fig_5}(2a,2b) show the absolute values of the normalized Fourier components $\bar{Z_n} = |Z_n|/\sum_{n=0}^{\infty}|Z_n|$ in the cases (i) and (ii), respectively.
It can be seen that the PSF $\bar{Z_n}$ in the case (ii) has larger values of the normalized high-frequency Fourier components than in the case (i), which leads to the larger improvement factor by the optimization in the case (ii) than in the case (i).

\section{Discussion}

We considered two types of optimization problems for the entrainment of a quantum nonlinear oscillator to a harmonic drive with a periodic amplitude modulation in the semiclassical regime.
We derived the optimal waveforms of the periodic amplitude modulation by applying the optimization methods originally formulated for classical limit-cycle oscillators to the semiclassical phase equation describing a quantum nonlinear oscillator.
Numerical simulations for the quantum vdP oscillator with squeezing and Kerr effects showed that the optimization of the entrainment stability leads to visibly faster convergence to the entrained state than the simple sinusoidal waveform, while the optimization for the phase coherence provides only tiny enhancement of the phase coherence from the sinusoidal case.
These results were explained from the Fourier-spectral properties of the PSF.
The squeezing and Kerr effects induced asymmetry of the limit-cycle orbit in the classical limit and yielded PSFs with stronger high-harmonic components, resulting in larger optimization performance.
It was also shown that optimization provides better performance when the PSF of the limit cycle has stronger high-frequency Fourier components in both problems.

The optimal waveforms for three typical optimization problems, i.e., improvement of entrainment stability~\cite{zlotnik2013optimal}, phase coherence~\cite{pikovsky2015maximizing}, and locking range~\cite{harada2010optimal} (not considered in this study),
which have been discussed for classical nonlinear oscillators in the literature, are proportional to the differential of the PSF, integral of PSF, and PSF itself, respectively.
All these waveforms yield negative feedback to the phase difference between the oscillator and the periodic forcing.
It is interesting to note that these relations between the optimal waveforms and PSFs bear some similarity to the proportional-integral-differential (PID) controller in the feedback control theory; in the framework of the PID control for linear time invariant systems~\cite{aastrom1995pid}, the differential control is often used for improving convergence, the integral control is used for improving the steady-state property, and the proportional control is used for improving the stability of the system.
Thus, similar to the PID controller, combined use of the three types of optimization methods for nonlinear oscillators could yield even better performance for achieving specific control goals of entrainment.

Though we have considered only the optimization problems for the stability and phase coherence of the entrained state in the present study, we would also be able to apply other optimization and control methods developed for classical limit-cycle oscillators, e.g. the phase-selective entrainment of oscillators~\cite{zlotnik2016phase}
and maximization of the linear stability of mutual synchronization between two oscillators~\cite{shirasaka2017optimizing, watanabe2019optimization}, to quantum nonlinear oscillators by using the phase equation for a quantum nonlinear dissipative oscillator
under the semiclassical approximation.
Such methods of optimal entrainment could be 
physically implemented with semiconductor optical cavities~\cite{hamerly2015optical} or 
optomechanical systems consisting of optical cavities and mechanical devices~\cite{amitai2017synchronization}
exhibiting limit-cycle behaviors, and
useful in future applications of quantum synchronization phenomena in quantum technologies.

\begin{acknowledgments}
	The authors gratefully acknowledge stimulating discussions with N. Yamamoto.
	This research was financially supported by the JSPS KAKENHI Grant Numbers JP16K13847, JP17H03279, 18K03471, and JP18H03287, and JST CREST JPMJCR1913.
\end{acknowledgments}

\appendix

\section{Derivation of the optimal waveforms}

In this Appendix, we give the derivation of the optimal waveforms.
The optimization problems for the improvement of entrainment stability and enhancement of phase coherence are
rewritten as
\begin{align}
\label{opt_obj_fe2}
\mbox{maximize}~ \int_{0}^{2\pi} \left( -Z_x'(\theta) \right) E(\theta) d\theta, ~ \mbox{s.t.}~\la E^2 (\theta)\ra_{\theta} = P,
\end{align}
and
\begin{align}
\label{opt_obj_mc2}
\mbox{maximize}~ \int_{0}^{2 \pi} 
\left( - \int_{\theta}^{\theta + \Delta_{\psi}} Z_x(\phi) d\phi \right) E(\theta) d\theta, ~ \mbox{s.t.}~\la E^2 (\theta)\ra_{\theta} = P,
\end{align}
respectively, where we assume $\psi^* = 0$ without loss of generality.
In order to analyze both problems together, we consider a general form of an optimization problem,
\begin{align}
\label{opt_obj_ge}
\mbox{maximize}~ \int_{0}^{2 \pi} g(\theta) E(\theta) d\theta, ~ \mbox{s.t.}~\la E^2 (\theta)\ra_{\theta} &= P,
\end{align}
where $g(\theta) = -Z_x'(\theta)$ for the entrainment stability 
and $g(\theta) = -\int_{\theta}^{\theta + \Delta_{\psi}} Z_x(\phi) d\phi$ 
for the phase coherence.

We consider an objective function
\begin{align}
S\{ E, \lambda \}
= \la g(\theta) E(\theta) \ra_\theta + \lambda \left( \la E(\theta)^2 \ra_\theta - P \right),
\end{align}
where $\lambda$ is a Lagrange multiplier. Then the extremum conditions are given by
\begin{align}
\frac{\delta S}{\delta E} = \frac{1}{2\pi} g(\theta) + \frac{\lambda}{\pi} E(\theta) = 0,
\end{align}
\begin{align}
\frac{\partial S}{\partial \lambda} = \la E(\theta)^2 \ra_\theta - P = 0.
\end{align}
The optimal periodic modulation is given by
\begin{align}
E(\theta) = -\frac{g(\theta)}{2\lambda} 
\end{align}
and the constraint is 
\begin{align}
\frac{1}{4 \lambda^2} \la g(\theta)^2 \ra_\theta = P,
\end{align}
which yields
\begin{align}
\lambda = -\sqrt{ \frac{1}{4 P} \la  g(\theta)^2 \ra_\theta },
\end{align}
where the negative sign should be taken in order that the maximized objective function becomes positive.

Therefore, the optimal periodic modulation is given by
\begin{align}
E(\theta) =
\sqrt{\frac{P}{ \la g(\theta) ^2 \ra_\theta }} g(\theta).
\end{align}
From the above result, the optimal waveform for the entrainment stability
is given by 
\begin{align}
E(\theta) =
-\sqrt{\frac{P}{ \la Z_x'(\theta) ^2 \ra_\theta }} Z_x'(\theta)
\end{align}
and that for the phase coherence is given by 
\begin{align}
E(\theta) =
-\sqrt{\frac{P}{ \la ( \int_{\theta}^{\theta + \Delta_{\psi}} Z_x(\phi) d\phi )^2 \ra_\theta }} \int_{\theta}^{\theta + \Delta_{\psi}} Z_x(\phi) d\phi.
\end{align}

When the limit cycle is symmetric and the $x$ component $Z_x$ of the PSF has a sinusoidal form, the optimal waveform is also given by a trivial sinusoidal function, because the differential and integral of a sinusoidal function are also sinusoidal.

%

\end{document}